\documentclass[preprint,12pt]{elsarticle}

\usepackage{amssymb}
\usepackage{amsmath}
\usepackage{moreverb,url}

\usepackage[colorlinks,bookmarksopen,bookmarksnumbered,citecolor=red,urlcolor=red]{hyperref}
\usepackage{float, caption}
\usepackage{subcaption}
\usepackage{multicol}
\journal{Journal of Cycling and Micromobility}
\begin{document}

\begin{frontmatter}

\title{Comparing E-bike and Conventional Bicycle Use Patterns in a Public Bike Share System: A Case Study of Richmond, VA}

\author[inst1]{Yifan Yang}
\author[inst1]{Elliott Sloate}
\author[inst2]{Nashid Khadem}
\author[inst2]{Celeste Chavis, Ph.D., P.E.}
\author[inst3]{Vanessa Fr\'ias-Mart\'inez, Ph.D.}

\affiliation[inst1]{organization={Computer Science Department, University of Maryland},%Department and Organization
            addressline={8125 Paint Branch Dr}, 
            city={College Park},
            postcode={20742}, 
            state={MD},
            country={United States}}

\affiliation[inst3]{organization={College of Information Studies and UMIACS, University of Maryland},%Department and Organization
            addressline={8125 Paint Branch Dr}, 
            city={College Park},
            postcode={20742}, 
            state={MD},
            country={United States}}

\affiliation[inst2]{organization={Department of Transportation \& Urban Infrastructure Studies, Morgan State University},%Department and Organization
            addressline={1700 E. Cold Spring Ln.}, 
            city={Baltimore},
            postcode={21251}, 
            state={MD},
            country={United States}}

\begin{abstract}

\textcolor{black}{Bicycle-sharing systems have emerged as a viable transportation alternative in numerous urban areas, owing to their multifaceted benefits. These benefits include reduced transportation expenses, health improvements, and decreased emission levels. While extensive research has been conducted on travel behaviors in shared bicycle systems, there is currently a lack of research on travel behaviors changes with the introduction of e-bikes. 
This study presents a comprehensive analysis of the similarities and differences between e-bike (pedelec) and conventional bicycle use in a bike share system in Richmond City, Virginia.}

\textcolor{black}{The results show that pedelecs are generally associated with longer trip distances, shorter trip times, higher speeds, and lower rates of uphill elevation change. The origin-destination analysis considering the business, mixed use, residential, and other uses shows extremely similar trends, with a large number of trips staying within either business or residential locations or mixed use. The roadway use analysis shows that pedelecs are used farther outside of the city than bikes. }

\end{abstract}


%%Research highlights
\begin{highlights}
\item Comprehensive analysis between pedelec and bicycle use in Richmond, Virginia.
\item Pedelecs are associated with longer trip distances, shorter trips times and lower rates of uphill elevation change.
\item Pedelecs are associated with higher average numbers of trips on major roads than bikes.
\item Pedelecs origin-destination analysis shows associations with mostly exercise (green areas) and recreational activities.

\end{highlights}

\end{frontmatter}

\section{Introduction}

The bicycle has become a legitimate transportation option in many cities due to its many benefits. Lower transportation costs, health improvement, and lower emission rates are some critical benefits of a bike ride. In congested cities, cycling is an efficient mode of transportation. Global climate change and energy security concerns are also growing, reflected in the sustainable transport system. Bike sharing – a service in which bikes are made available to the public, sometimes for a fee – is growing worldwide to keep pace with these growing concerns. Public bike share programs offer a solution to short trips and, through integration with public transit, serve as a first- and last-mile solution. People consider bike share a greener and better way of life \citep{yan2017rental}. However, due to users' age and different health conditions, and local areas' infrastructure and terrain, some people cannot regularly use a bicycle. Electrically assisted bikes (e-bikes) are being introduced in many bike share systems to overcome these issues. The introduction of e-bikes has reduced traditional all-human powered cycling barriers, including the perception of fitness needed, age, terrain condition, and travel speed \citep{rose2012bikes, macarthur2013bikes, dill2012electric, gordon2013experiences}.

A large body of research exists on travel behavior in general, and specific to bike share systems. GPS data provides researchers with the opportunity to analyze travel behavior as a function of built environment characteristics
and with respect to convenience, safety, and leisure \citep{chen2018gps}. 
Several studies have found that cyclists prefer facilities on flat, low-volume roads with slow traffic or separated bike infrastructure \citep{chen2018gps, richmond2014,vieira2010querying,mcneil2015influence,winters2010route}. This research has been used to develop level of traffic stress measurements \citep{sorton1994bicycle, klobucar2007network, 2012mobilizing, wu2018predicting}, determine the location of bicycle infrastructure \citep{nickkar2018bicycle, garcia2012optimizing, khadem2019bike, banerjee2020optimal}, and provide route guidance \citep{wu2018predicting, caggiani2017real, schuijbroek2017inventory} as a function of traffic volumes, speeds and bike infrastructure provision.

Current research is lacking on travel behavior changes with the introduction of e-bikes in bike share systems. As shown earlier, e-bikes remove some biking barriers associated with health and physical ability. Physical ability has also been linked to route choice factors, such as route length and terrain \citep{lee2016system}. Additionally, e-bikes may influence safety-related factors such as traffic speed and perceived safety at stops \citep{steer}. By allowing quicker acceleration and reducing the speed differential between bikes and vehicles at upgrades, e-bikes may influence travel behaviors. 

\textcolor{black}{As bike share systems continue to incorporate e-bikes, it becomes critical to understand whether the introduction of e-bikes in shared bike systems are changing cycling behaviors, and if so, in what manner. Hence, we carry out a comparative study of e-bikers’ travel behaviors with respect to conventional cyclists to explore the impact of e-bikes on cycling trip characteristics.} This paper advances the state of the art by analyzing the differences between pedelec and conventional bicycle trips within a bike shared system in terms of trip length, duration, speed and elevation; the types of trip origin and destinations and the types of roads used.

Specifically, we focus in the RVA bike share system of Richmond, Virginia, and use their RVA Bike Share data sources to carry out the analysis. 
The paper is organized as follows. We first describe the state of the art (Section 2), followed by a description of the study site, the dataset we use in our analysis, and the methodology followed to answer the proposed research question (Section 3). We present our main findings in Section 4, and finalize with conclusions in Section 5.

\section{Literature Review}

\subsection{Bike Share Programs}
Introduced by transportation planners and often called rental bikes or public use bicycle programs, bike share programs have been implemented worldwide \citep{schuijbroek2017inventory}. \textcolor{black}{There exist many bike sharing systems, with public bicycle sharing and recreational bicycle-sharing systems being the most common; including universities that have introduced bike share programs exclusive to their students for commuting on campus. The majority of public bicycle sharing systems in urban settings operate bicycle rental programs that aim to give commuters an accessible and time-efficient transportation mode in congested areas. }
%Different business groups (Bewegen, Copr, CycleHop, Citi Bike, Lime bike, and many more) operate bicycle rental programs. 
Users can rent both docked and dockless bicycles depending on their origin and destinations and bike share companies’ systems. People can also rent for a few hours or for a few days, depending on their needs. Bike-sharing programs allow participants to use a bicycle as needed without bicycle ownership costs and responsibilities \citep{shaheen2010bikesharing}.

Third-generation bike share programs have greatly minimized issues of theft. Bike-sharing applications now use different technologies, including smartphone use, GPS tracking, debit/credit card payment systems, real-time bike inventories, and many more to track the bike and user’s route to prevent theft, creating an incentive to bring the bikes back promptly \citep{bryant2013finding}. More than 1,000 cities have a bike sharing program, and the numbers are increasing. In China, Hangzhou has the world’s most extensive bike-sharing program, well integrated with other public transport forms. In the USA, bike sharing is also increasing in popularity, with both docked and dockless bikes shares popular among rider groups. In 2016, the total number of bike share bikes was 42,500, which doubled in 2017 to 100,000 bikes. In 2017, dockless bike share companies added almost 44,000 bikes countrywide (U.S), while 14,000 station-based bikes were added to the system \citep{nacto}. 

\subsection{Introduction of E-bikes}

\textcolor{black}{An electric bike or e-bike is a bicycle that has an electric motor that provides power assistance up to speeds of 25 km/hour. This kind of bike is engaged with a throttle grip or pedaling, and power can only be engaged by pedaling, also called pedal-assist or pedelec \citep{rose2012bikes}. E-bikes are an excellent addition to bike share programs since they reduce many barriers to pedal cycling such as age, health determinants, steep terrain or lack of time, %and end-of-tour facilities 
\citep{rose2012bikes, macarthur2013bikes, dill2012electric, gordon2013experiences}.  
An e-bike is also faster than a conventional bicycle and enables users to take longer trips, even on hilly routes. E-bikes can also replace car or bus trips and avoid rush hour traffic by offering competitive travel speeds. 
Most e-bikes look similar to bicycles and have their battery pack fitted in a different location such as the seat post, bike frame, or rear rack \citep{johnson2015extending}. Though the power assistance makes the riding more comfortable, users still need to pedal, which provides physical activity benefits \citep{louis2012electrically}. }
%E-bikes are attractive to people with injuries, or those who are less fit or older.

\textcolor{black}{Due to the many benefits, e-bikes are becoming more common in different countries. For example, in Europe, there has been a significant increase in e-bikes sales \citep{langford2013north}, and e-bikes account for between 12\% to 15\% of total bicycle sales \citep{fishman2016bikes, chavanon_2021}. }
%Several studies have shown that access to e-bikes increases the volume of trips as well as the distance traveled \citep{wu2012red, fyhri2015effects}. } 
E-bikes are also energy efficient and environmentally preferred modes compared to other motorized transportation modes \citep{khatri2016modeling}.

\subsection{Impacts on Travel Behavior}
%\ and Mode Choice}

 \textcolor{black}{Introducing e-bikes in any city can influence the travel behavior of the cyclists. Next, we describe the main findings of several studies that have used primary data (GPS) as well as survey-based data to analyze travel behavior across e-bikes and conventional bicycles.}
 %and mode choice. 

%\\ \textbf{Primary Data studies:} 
\textcolor{black}{A Norwegian study randomly selected 66 individuals and gave them e-bikes to use during a limited period of time. After that period e-bike usage was compared against a control group of 160 conventional cyclists \citep{fyhri2015effects}. The analysis showed significant changes between e-bike and conventional bicycle users with the former being associated to larger average number of daily trips (from 0.9 to 1.4); and to larger riding distances 
%that the number of cycling trips increased from 0.9 days to 1.4 days. The riding distance also increased 
(from 4.8 km to 10.5 km). The proportion of trips by e-bike, as a share of all transport, also increased from 28\% to 48\%.  In another study \citet{langford2013north}, the authors used e-bike share data from North America's first e-bikesharing system (cycleUshare), at the University of Tennessee, Knoxville, to assess travel behaviors and attitudes. 
The study revealed that speed and comfort play a vital role in selecting an e-bike instead of a conventional bike. The authors also observed that the bike share system expanded user mobility and diversified trip purposes. The study also revealed that e-bike riders rode 13\% farther than their conventional bicycle counterpart. 
%Some other studies also show that e-bike users travel a greater distance than traditional bicycle users\textcolor{red}{we need reference here}. 
Another study by \citet{cherry2016dynamics} found that the average trip distance traveled by e-bike increased 4 km between 2006 to 2012. Simiarly, a study carried out in two Chinese cities showed that the use of e-bikes increased the distance traveled 
%also improved the vehicle kilometer traveled (VKT), 
by 9\% and 22\% in Shanghai and Kumming, respectively. The travel speed observed was also higher for e-bikes than traditional bikes, 15\% in Shanghai and 10\% in Kumming \citep{cherry2007use}. }
%\textbf{Survey-based studies:} 
\textcolor{black}{Finally, a survey-based study carried out by \citet{macarthur2013bikes} focused on answering two questions: whether e-bikes would increase the number of trips people make and whether it would increase riding frequency. 
%get more people to ride and will e-bikes increase riding frequency?
The authors found that e-bikes may increase cycling participation, and that almost 55\% of people would start riding daily and 93\% weekly, after getting e-bikes. }

Although previous studies have analyzed pedelec use, our paper advances the state of the art by looking into a comprehensive analysis of the 
differences between pedelec and conventional bicycle trips within Richmond's bike share system (RVA) in terms of trip length, duration, speed and elevation; the types of trip origin and destinations and the types of roads used.

\begin{table}[]
\caption{RVA Bike Share standard membership categories.}
\centering
\begin{tabular}{lp{6.5cm}lll}
Membership & Description                                   & Current Price \\\hline
Annual                        & Unlimited   45-min rides for 1 year. 1 bike per membership.                                                   & \$96                         \\
Monthly                            & Unlimited   45-min rides for 1 month. 1 bike per membership.                                                  & \$18                         \\
Weekly Pass                        & Unlimited   45-min rides for 7 days. One bike per pass, possible to purchase up to 4   passes at the kiosk.   & \$12                        \\
Day Pass                           & Unlimited 45-min   rides for 24 hours. One bike per pass, possible to purchase up to 4 passes at   the kiosk. & \$6                       \\
Go Pass                            & Receive a   pass to unlock bikes but pay per 45-min ride. Not available at kiosk.                             & \$1.75 per   ride           \\
One-way Trip   Pass                & One 45-min   ride. A pass is dispensed at the kiosk to unlock the bike. May rent up 4   bikes at once.        & \$1.75 per   ride          
\end{tabular}

\label{tab:1}
\end{table}

\section{Methods and Material}
\subsection{Study Site: RVA Bike Share}
In 2017, Richmond, Virginia, launched the RVA Bike Share, a dock-based bike-sharing system. At its launch, the system offered only conventional pedal bikes. Beginning in March 2019, RVA Bike Share began converting the traditional bikes to e-bikes, also available in the docked system. At the time of this study, a total of 220 bikes (both traditional and pedelec) were available across 19 stations throughout central Richmond, Virginia;
and the Main Street Station was not yet in operation.

There are six general pricing structures for the RVA Bike Share (see Table \ref{tab:1} for summary). Bike share trips can be charged per trip (Go Pass and One-way Trip Pass) or people can pay to take an unlimited number of trips within a year, month, week, or day (Annual, Monthly, Weekly and Day passes). Annual members include Annual Founding members, who were early adopters and who currently share pricing structure with annual members. Trips over 45 minutes are subject to average fees of \$3 per 30 minutes. Figure \ref{fig:1} shows the locations of the RVA Bike Share stations in relation to the bicycle facilities. During the study period two special memberships were offered. The Fall Offer Pass was offered from October to December and the RVA Mural Bike Tour occurred in August.

 \iffalse
\begin{figure}
\centering
\begin{subfigure}{.5\textwidth}
  \centering
  \includegraphics[width=.8\textwidth]{figures/figure1.png}
  \caption{RVA Bike Share system.}
  \label{fig:1}
\end{subfigure}%
\begin{subfigure}{.5\textwidth}
  \centering
  \includegraphics[width=\textwidth]{figures/figure2.png}
  \caption{Total number of trips to and from station (min = 3 trips, max = 1208).}
  \label{fig:2}
\end{subfigure}
\caption{Map of RVA Bike Share system.}
\end{figure}
\fi

\begin{figure}[!ht]
   \centering
   \includegraphics[width=0.6\textwidth]{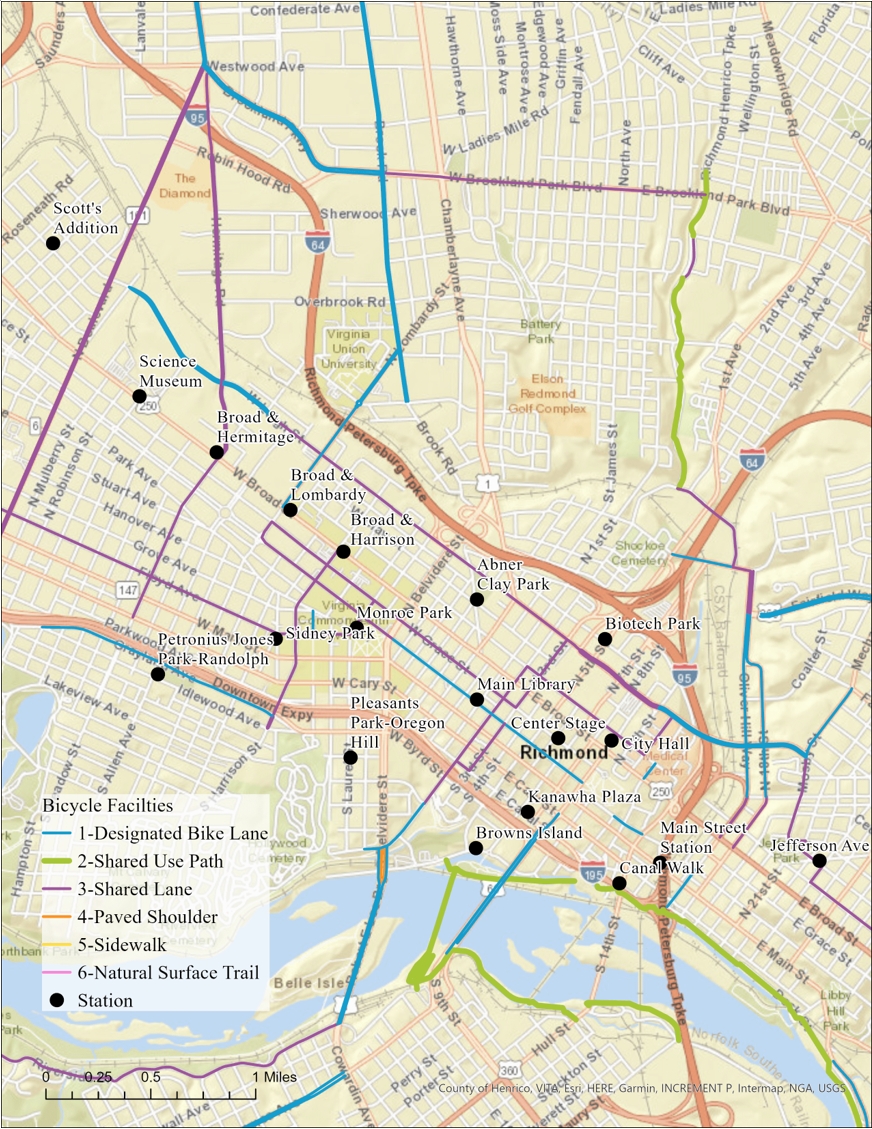}
   \caption{ RVA Bike Share system. \textcolor{black}{The basemap was generated with Esri ArcGIS and the following sources: Esri, HERE, Garmin, GeoTechnologies, Inc., USGS and EPA.}}
   \label{fig:1}
 \end{figure}

 \subsection{Trip Data}
RVA Bike Share data was received from the bike share operator Bewegen. The dataset contained a total of 4,075 trips collected during the first week of each month from April 2019 (the beginning of the transition to pedelec bikes) to December 2019 (the last full month of data at the start of the study). 
\textcolor{black}{Only one week per month was shared by the operator due to their lack of capacity to pre-process longer periods of time. Nevertheless, when selecting the week each month, we ensured that the week was normal and free of major operational and weather events.}
A final total of 17,553 trips were recorded from April to December 2019.

The data contained the following information:

\begin{multicols}{2}
\begin{itemize}
\setlength\itemsep{0em}
\item	Bike unlock date
\item	Bike unlock time 
\item	Bike lock date
\item	Bike lock time
\item	Membership type
\item	Distance (miles)
\item	Duration (minutes)
\item   Speed (kph)
\item	Bike ID
\item	Bike Type (bike or pedelec)
\item	Cost of trip 
\item	Start station
\item	End station
%\item	Route ID 
\item	User ID
\end{itemize}
\end{multicols}
 
Trips shorter than 30 seconds, longer than 3 hours and trips that covered more than 100 km (62 mi) were filtered out to eliminate outliers from the dataset that could be related to incorrect system performance or to people simply trying out the bicycles but not traveling. The total final number of trips was 3,519 with 2,257 pedelec trips and 1,262 traditional bicycle trips. Figure \ref{fig:2} shows the total number of trips that started and ended at each station, and Table \ref{tab:2} presents the number of trips which started and ended at each station during the study period.

 \begin{figure}[!ht]
   \centering
  \includegraphics[width=\textwidth]{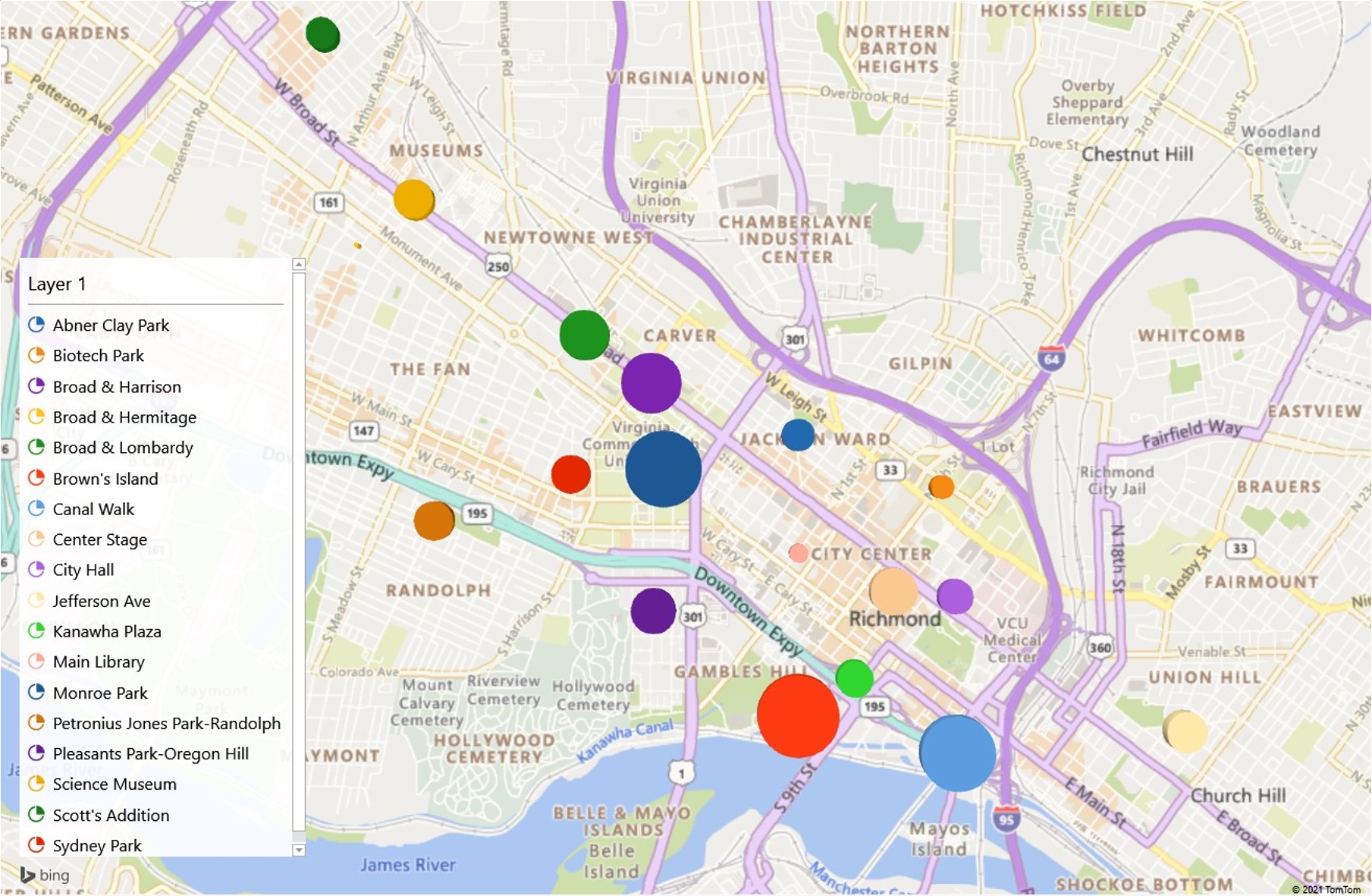}
  \caption{Total number of trips to and from station (min = 3 trips, max = 1208). \textcolor{black}{The basemap was generated with Bing Maps and with TomTom as its basemap data source.}}
  \label{fig:2}
 \end{figure}

\begin{table}[]
\centering
\caption{Number of trips originating and destined to each station.}
\begin{tabular}{llll}
 
Station                         & Origin & Destination & Total \\\hline

Abner Clay Park                 & 120    & 111         & 231   \\
Biotech Park                    & 34     & 40          & 74    \\

Broad \&   Harrison             & 308    & 258         & 566   \\
Broad \& Hermitage              & 0      & 3           & 3     \\

Broad \&   Lombardy             & 216    & 218         & 434   \\
Brown's Island                  & 569    & 639         & 1208  \\

Canal Walk                      & 455    & 466         & 921   \\
Center Stage                    & 210    & 176         & 386   \\

City Hall                       & 93     & 73          & 166   \\
Downtown YMCA                   & 0      & 1           & \textcolor{black}{1}   \\

Jefferson Ave                   & 132    & 134         & \textcolor{black}{266}    \\
Kanawha Plaza                   & 113    & 106         & \textcolor{black}{219}    \\

Main Library                    & 39     & 47          & \textcolor{black}{86}    \\
Monroe Park                     & 563    & 602         & \textcolor{black}{1165}    \\
 
Petronius Jones   Park-Randolph & 122    & 137         & \textcolor{black}{259}    \\
Pleasants Park-Oregon Hill      & 177    & 148         & \textcolor{black}{325}    \\
 
Science Museum                  & 129    & 104         & \textcolor{black}{233}    \\
Scott's Addition                & 81     & 83          & \textcolor{black}{164}    \\

Sydney Park                     & 158    & 173         & \textcolor{black}{331}   
\end{tabular}
\label{tab:2}
\end{table}

\subsection{Data Summary}

\begin{figure}
\centering
\begin{subfigure}{.49\textwidth}
  \centering
  \includegraphics[width=\textwidth,trim=4 4 1 4,clip]{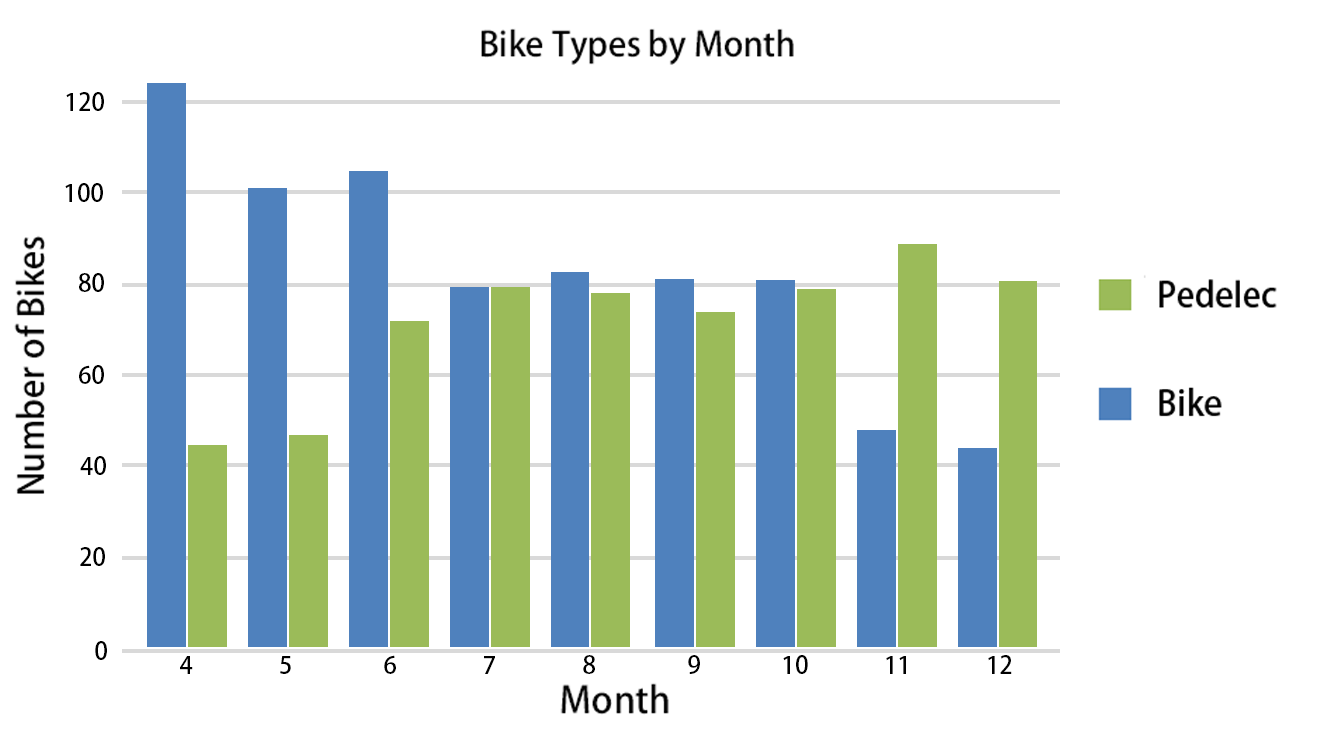}
  \caption{Number of bikes and pedelecs available by month.}
  \label{fig:13}
\end{subfigure}
\begin{subfigure}{.49\textwidth}
  \centering
  \includegraphics[width=\textwidth,trim=4 4 1 4,clip]{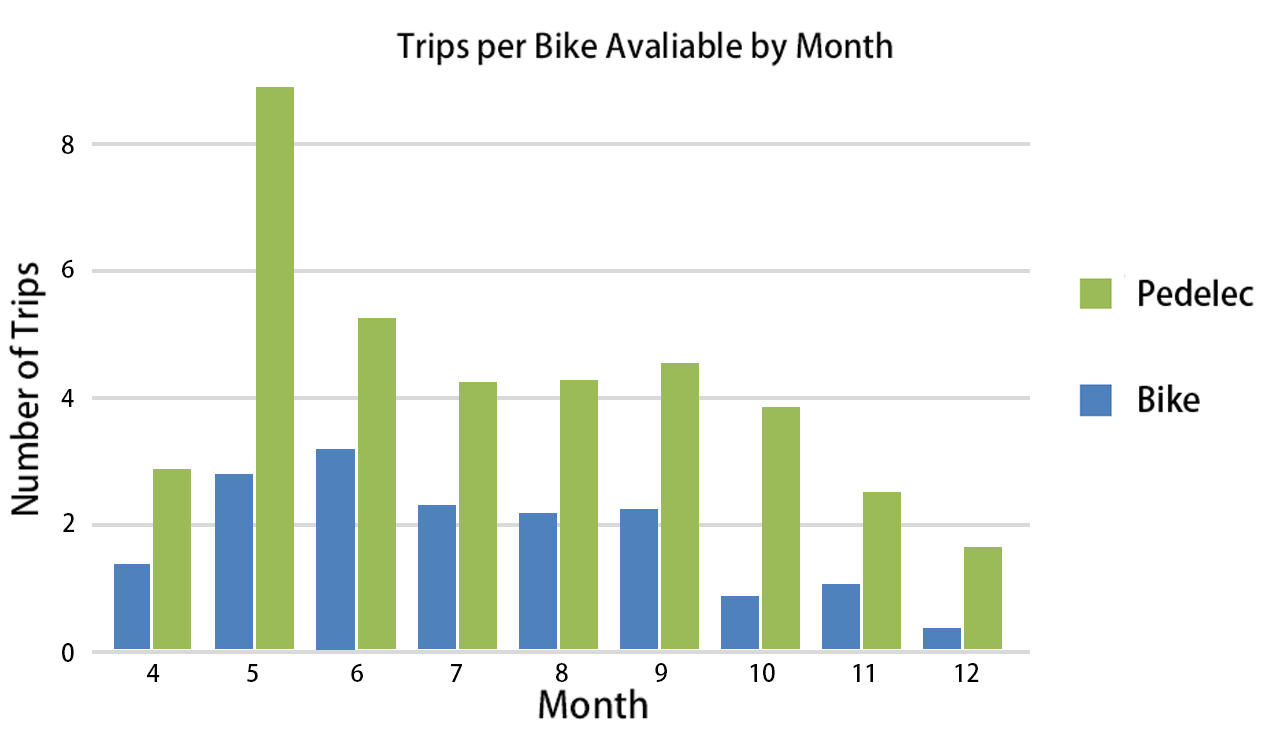}
  \caption{Number of trips made per available bike or pedelec by month}% extracted from the first week of each month.}
  \label{fig:12}
\end{subfigure}
\begin{subfigure}{.49\textwidth}
  \centering
  \includegraphics[width=\textwidth,trim=4 4 1 4,clip]{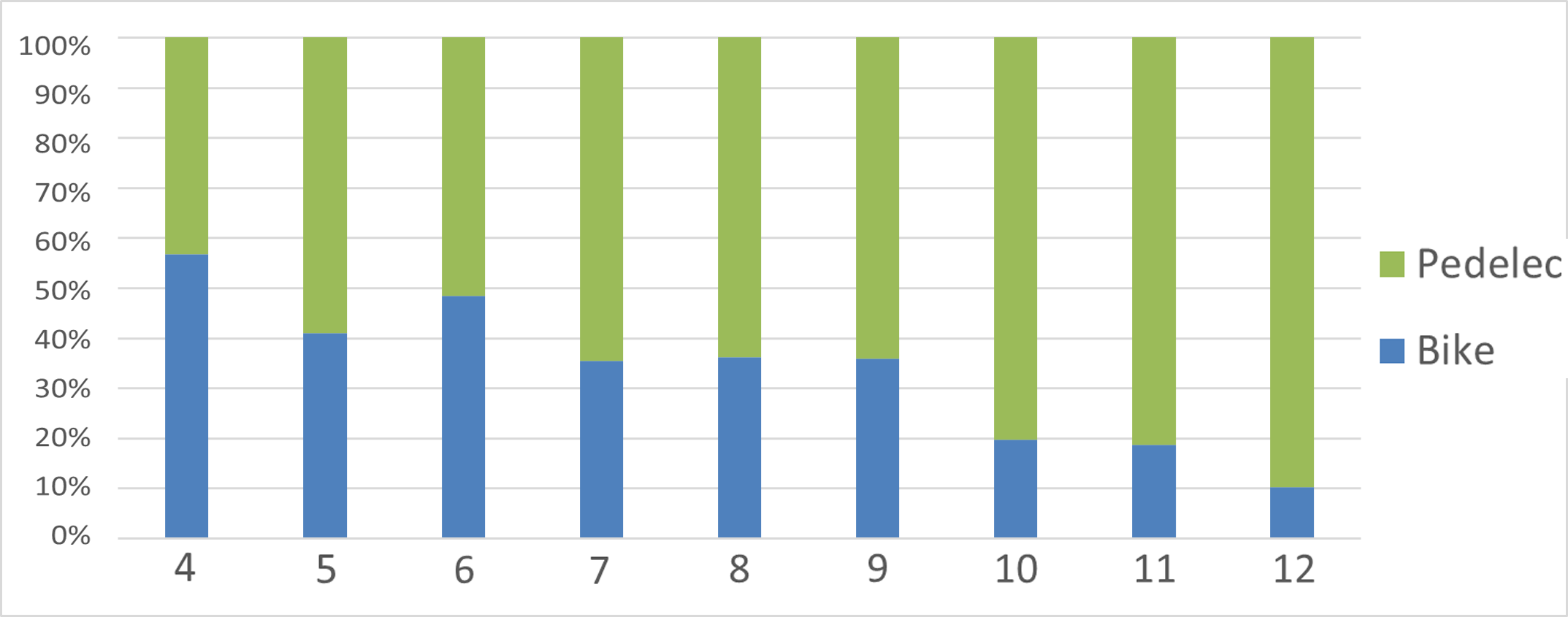}
  \caption{Percentage of trips by bike type per month.}% in 2019.}
  \label{fig:3}
\end{subfigure}
\begin{subfigure}{.49\textwidth}
  \centering
  \includegraphics[width=\textwidth,trim=4 4 2 4,clip]{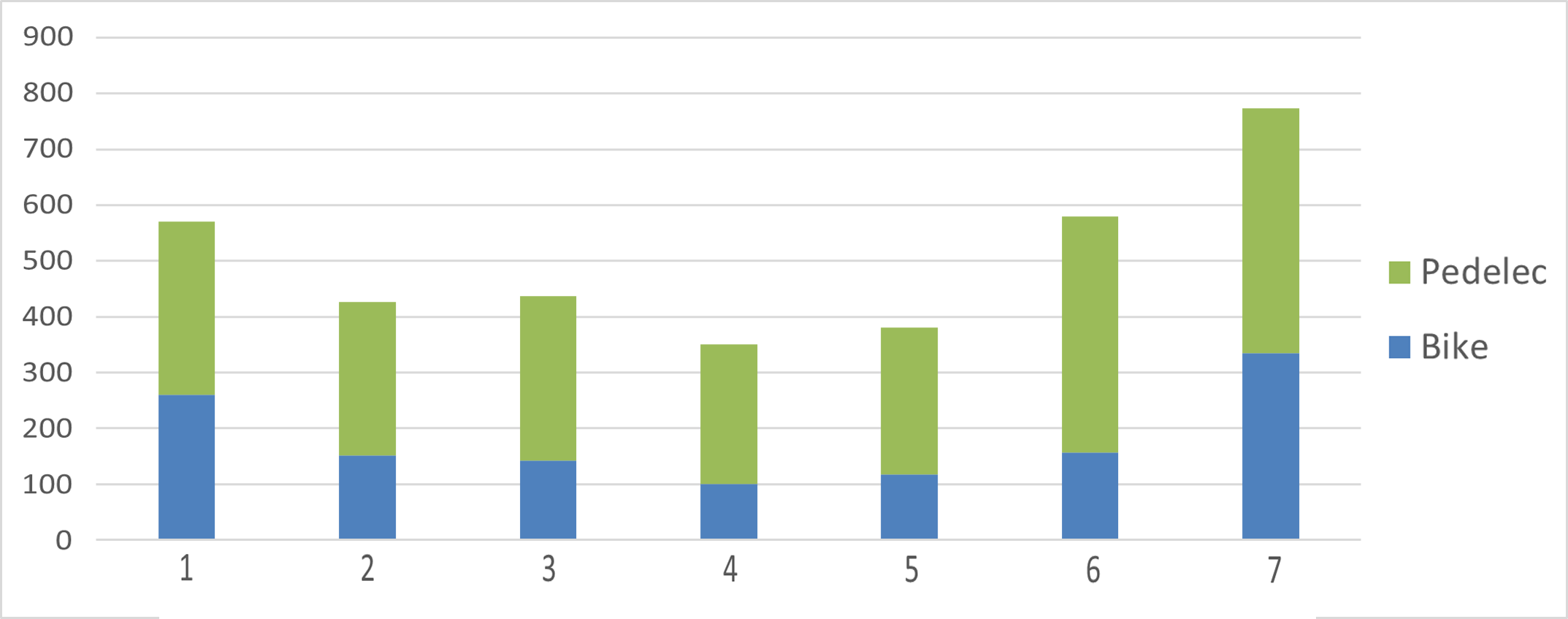}
  \caption{Total number of trips per day of week (1= Sunday, 7=Saturday).}
  \label{fig:4}
\end{subfigure}
\caption{\textcolor{black}{Bike and trip characteristics over the study period (first week of each month) segmented by bike type.}}
\label{fig:test}
\end{figure}

RVA Bike Share began to introduce pedelecs into the system in March 2019. As e-bikes were introduced gradually into the 19 docked stations; bikes were being removed from the fleet, although not necessarily with a 1:1 correspondence. The introduction of e-bikes and the removal of bikes was not homogeneous, neither by station nor by month; and the total number of e-bikes and bikes available changed on a monthly basis, as can be observed in Figure \ref{fig:13}. The percentage of trips by pedelec and the rate of trips made per available e-bike increased over time (see Figures \ref{fig:12} and \ref{fig:3}), even as e-bikes and bikes were available at similar rates (see months 7-10 in Figure \ref{fig:13}). 
However, it is not clear if the difference in pedelec use reflects differences in trip purposes or familiarity with the system since pedelecs are only identified by a lightning bolt on the back of the bike. 
By December 2020, the fleet of pedelec bikes was roughly 65\%, but over 90\% of trips were taken on a pedelec bike.

\begin{figure}%{\textwidth}
  \centering
  \includegraphics[width=\textwidth,trim=4 4 1 4,clip]{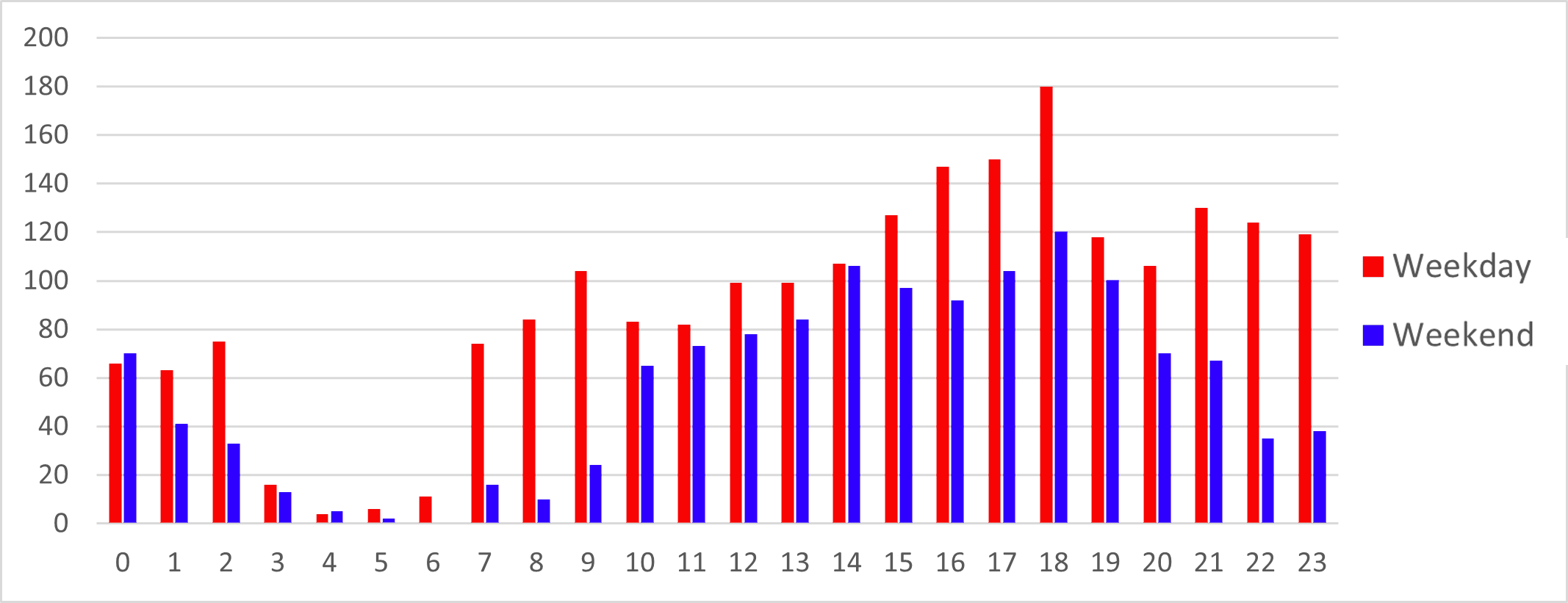}
  \caption{\textcolor{black}{Total number of weekday and weekend trips per hour of the day.}}
  \label{fig:5}
\end{figure}

Figure \ref{fig:4} shows the number of trips made \textcolor{black}{by day of the week.} The busiest days were Sunday (represented by =1) and Saturday (=7). More trips were made on Friday than other weekdays. Figure \ref{fig:5} shows the total number of trips made by hour of the day on the weekday and weekend. The busiest time period for RVA bike share on the weekend is between 2 pm and 7 pm, and on the weekdays, it is 4 pm to 6 pm, following workday peak trends. Morning bike share use was more prevalent on the weekdays, capturing people who use the bike share for commuting.

\begin{figure}
\centering
\begin{subfigure}{\textwidth}
  \centering
  \includegraphics[width=\textwidth,trim=4 4 1 4,clip]{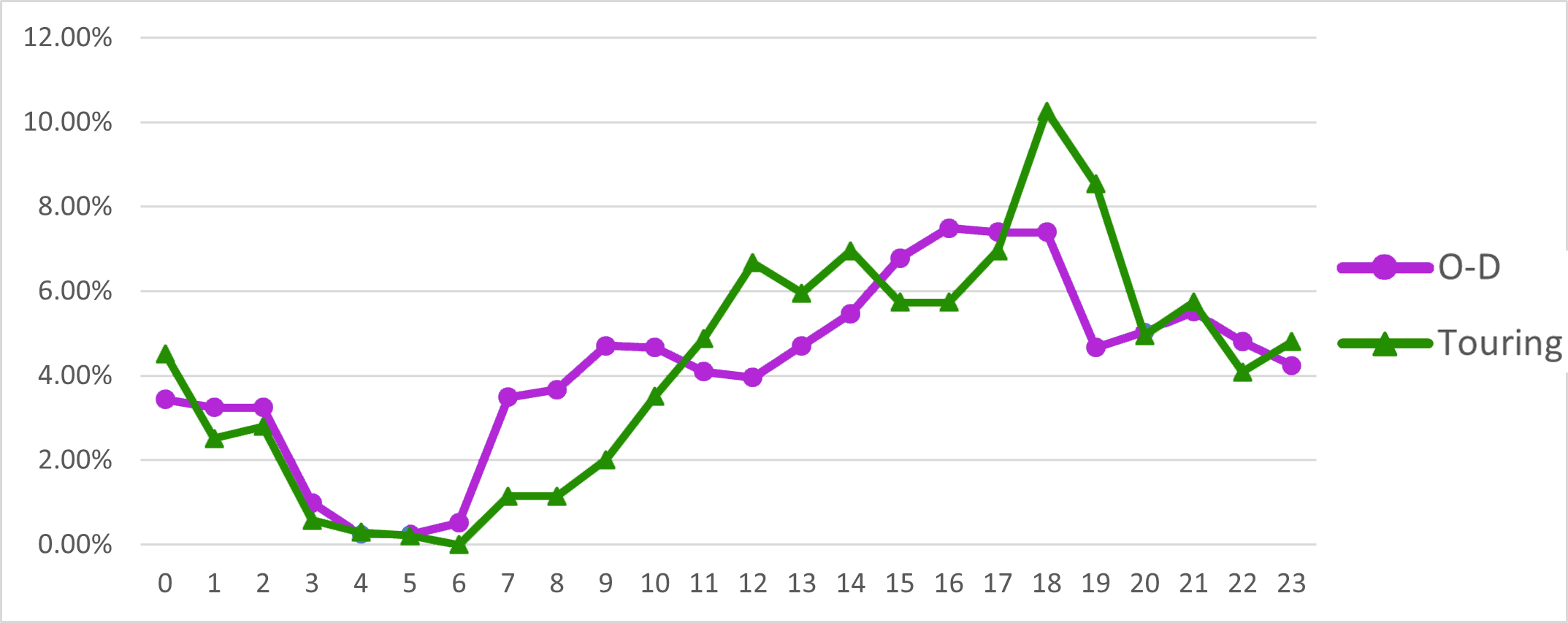}
  \caption{Percentage of trips by time of day.}
  \label{fig:6}
\end{subfigure}
\begin{subfigure}{\textwidth}
  \centering
  \includegraphics[width=\textwidth,trim=4 4 1 4,clip]{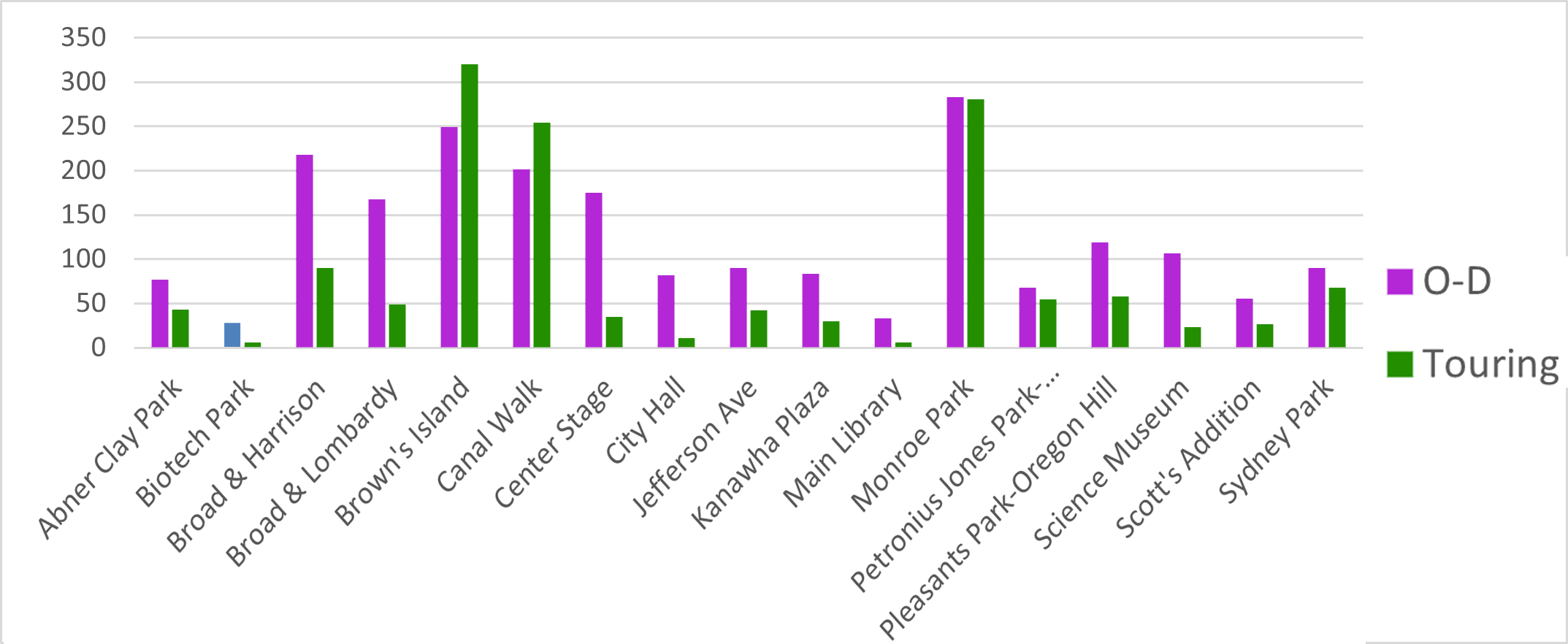}
  \caption{Number of trips originating from each station.}%: O-D vs. Touring Trips.}
  \label{fig:7}
\end{subfigure}
\caption{\textcolor{black}{Trip summary by trip type (touring vs. O-D).}}
\label{fig:test}
\end{figure}

Trips are categorized into two categories. Touring trips are trips that start and end at the same station while O-D trips have a different origin and destination station. Notice that both touring and O-D trips can include small stops along the way, and that both trip categories end when the (e-)bike is docked again at a station. As shown in Figure \ref{fig:6},  the morning trips are dominated by O-D trips, likely commuting trips. Figure \ref{fig:7} shows the total number of O-D and touring trips by the origin station. Brown’s Island, Canal Walk, and Monroe Park, which have recreational land use, have the highest number of trips. Several of the downtown stations, such as Brown \& Harrison and Center Stage had more O-D trips compared to touring trips.

\subsection{Membership Types}

\begin{figure}
\centering
\begin{subfigure}{\textwidth}
  \centering
  \includegraphics[width=\textwidth,trim=4 4 1 4,clip]{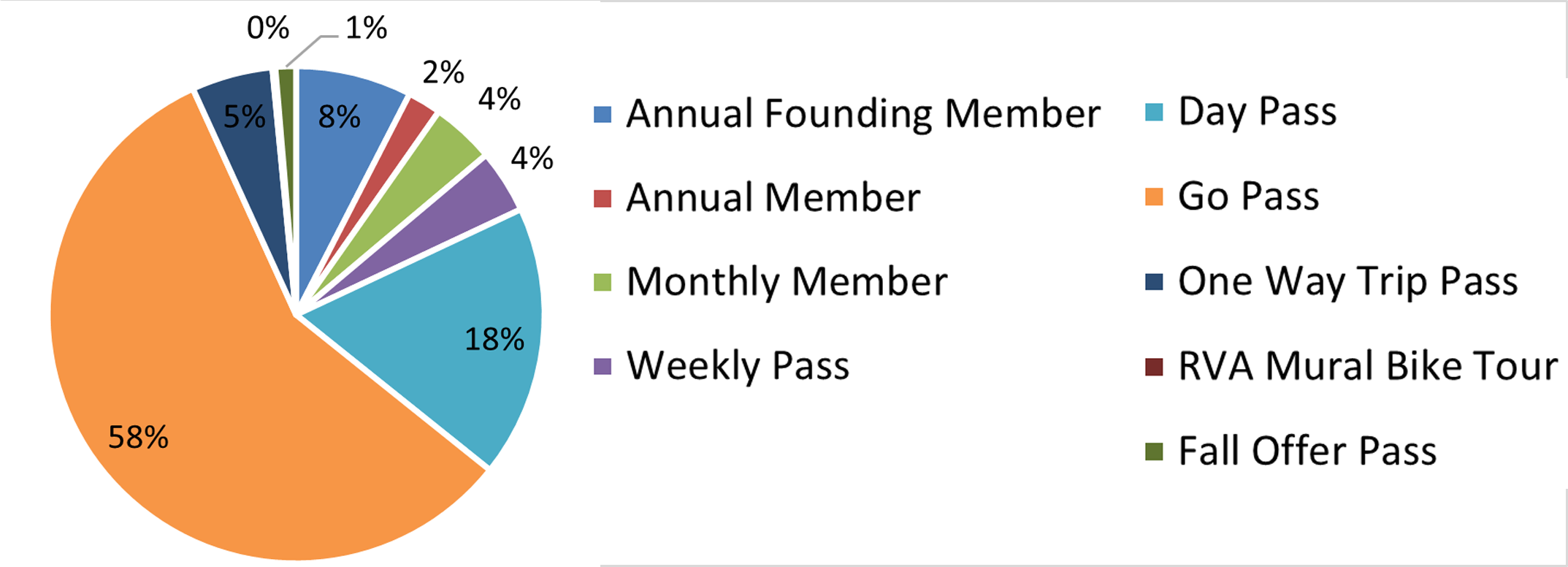}
  \caption{Percent of Trips.}% by Membership Type.}
  \label{fig:9}
\end{subfigure}
\begin{subfigure}{\textwidth}
  \centering
  \includegraphics[width=\textwidth,trim=4 4 2 4,clip]{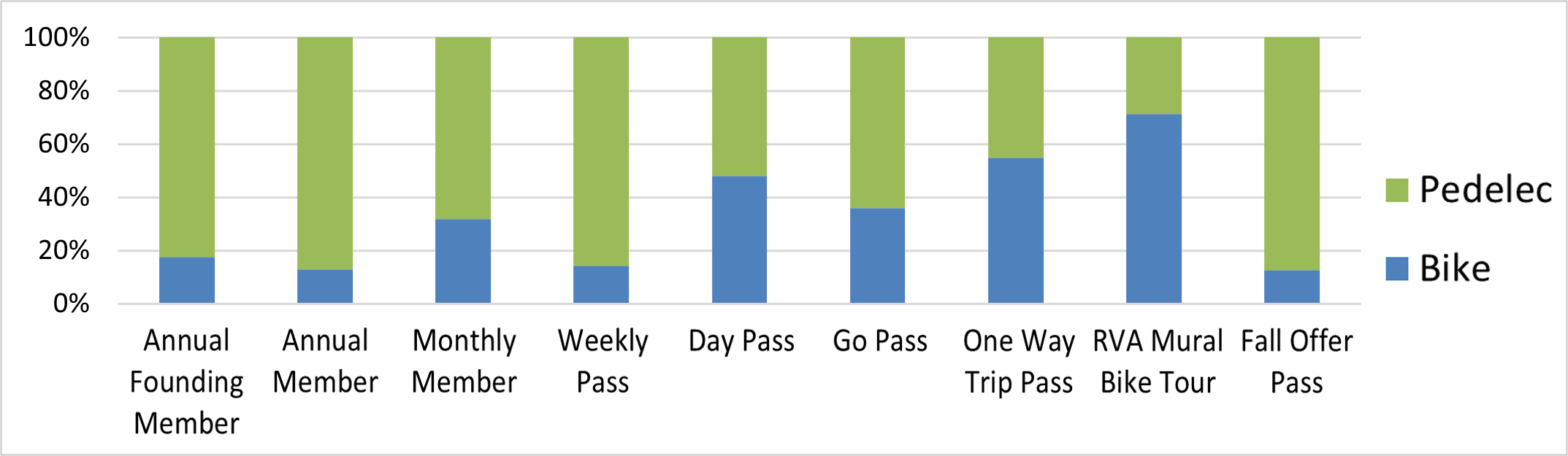}
  \caption{Percent of trips made by bike type.}% on Pedelecs and Bikes by Membership Type.}
  \label{fig:10}
\end{subfigure}
\begin{subfigure}{\textwidth}
  \centering
  \includegraphics[width=\textwidth,trim=4 4 1 4,clip]{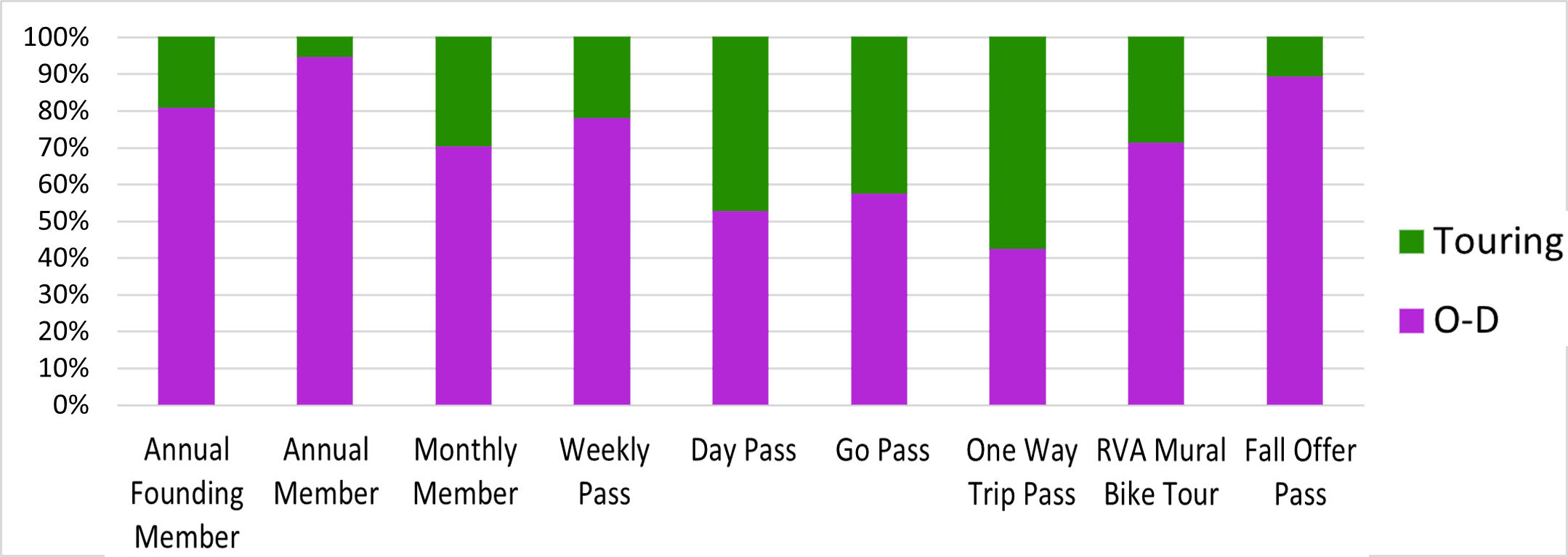}
  \caption{Percent of trips by trip type.}%Touring vs. O-D by Membership Type.}
  \label{fig:11}
\end{subfigure}%
\caption{Trip summaries by membership type.}
\label{fig:test}
\end{figure}

Over the study period, the majority of trips (75\%) were made by those who paid by the ride (go passes and one-way trip passes). Annual, monthly, and weekly membership trips were only 14\% of all trips; see Figure \ref{fig:9}. Subscription members had the highest rates of trips made by pedelecs; whereas one-way trip passes, day passes, and go passes had the lowest rate of pedelec use; see Figure \ref{fig:10}.

As shown in Figure \ref{fig:11}, annual members (and other subscription members) take more trips that have a different start and end station (O-D trips) while short-term members and pay per trip riders tend to take more trips that start and end at the same station (touring trips).

On average O-D trips were about 3.7 km and touring trips 5.5 km; see Table \ref{tab:3}. Annual founding members had the largest difference in the average touring kilometers (8.0 km) and RVA Mural Bike Tour had the largest O-D kilometers (10.6 km) followed by monthly members (4.2 km). The longest trips were done by RVA Mural Bike Tour (92 min) followed by monthly members (32.5 min). 
%RVA Mural %Bike Tour trips had the longest travel time at 92 minutes. 
On average, touring trips were about twice as long as O-D trips (45 mins vs. 24 mins).

\begin{table}[]
\centering
\caption{Average Trip Distance and Time Based on Membership Type, including standard and special membership types.}
\begin{tabular}{lllllll}
                         & \multicolumn{3}{l}{Average Distance (km)}              & \multicolumn{3}{l}{Average Time (min)} \\
Membership   Type        & O-D                      & Touring                  & All Trips                & O-D              & Touring             & All Trips             \\\hline
Annual Founding Member   & 3.2                      & 8.0                      & 4.2                      & 14.1             & 34.8                & 18.1                  \\
Annual Member            & 1.6                      & 1.6                      & 1.6                      & 7.6              & 14.6                & 8.0                   \\
Monthly Members          & 4.2                      & 5.5                      & 4.8                      & 32.5             & 48.2                & 39.9                  \\
Weekly Pass              & 3.7                      & 5.6                      & 4.5                      & 26.3             & 46.6                & 34.9                  \\
Day Pass                 & 1.8                      & 3.2                      & 2.3                      & 10.9             & 29.9                & 16.5                  \\
Go Pass                  & 3.7                      & 4.7                      & 4.3                     & 25.7             & 37.3                & 32.4                  \\
One Way Trip Pass        & 5.1                      & 6.0                      & 5.3                      & 25.4             & 35.6                & 27.6                  \\
RVA Mural Bike Tour      & 10.6                      & 1.3                      & 8.0                      & 92.0             & 18.2                & 70.9                  \\
Fall Offer Pass          & 1.6                      & 4.2                      & 2.7                      & 13.1             & 24.9                & 14.3                  \\
Grand Total              & 3.7                      & 5.5                      & 4.3                      & 24.4             & 44.8                & 32.5       
\end{tabular}
\label{tab:3}
\end{table}

\subsection{Methodology}
\textcolor{black}{Our objective is to analyze use differences between pedelecs and traditional bikes in the RVA bike share system of Richmond, Virginia. Specifically, we focus our analysis on the trip length, duration, speed and rate of elevation change; on the types of trip origin and destination and on the types of roads used for the period under analysis (nine months). }

We present our analysis in three broad threads: (1) spatio-temporal difference between pedelec and bicycle trips by looking into length, trip duration, speed and rate of elevation change; \textcolor{black}{(2) difference in terms of types of origin and destination locations visited during the cycling trips; } and (3) routing differences.

Retrieving the street segments associated with the GPS traces of a cycling trip is not straightforward since GPS sensors have errors, and more so in urban environments where, when surrounded by tall buildings, the GPS might lose the signal or record a location quite far away from the actual one. 

\textcolor{black}{The data for each trip consists of a list of GPS coordinates. We first use Mapbox’s Map Matching API to snap the GPS traces to actual road segments in the road network \citep{mapbox}. Mapbox’s Map Matching API takes lists of GPS coordinates as input, and outputs a matched list of coordinates of road segments.}
Internally, Mapbox uses the map-matching algorithm by Newson and Krumm, based on Hidden Markov Models (HMM), that finds the most likely street segment in the network that is represented by the collected GPS location \citep{newson2009hidden}.
\textcolor{black}{After retrieving the road segments, we query each segment in Open Street Maps (OSM) to identify the type of road.}

\textcolor{black}{To analyze the types of roads traveled, we assigned each street segment in a trip to a label characterizing the type of road. For analytical purposes, we considered three types of roads: major roads, minor roads and roads with cycleways.}\textcolor{black}{
We used the road typology definitions provided by Open Street Maps (OSM) to define these three types and defined as major roads roads with the OSM labels "motorway", "trunk", "primary", "secondary” and "tertiary"; as minor roads, road segments with the OSM labels “road”, “residential”, “living street”, “service”, and “pedestrian”; and, road segments with the OSM label “cycleway” were considered cycleways. }

\textcolor{black}{On the other hand, we compute the rate of elevation change for a trip as the quotient between the uphill elevation change for each segment in a trip divided by the trip duration, with the uphill elevation change calculated as the sum of the positive differences between the elevation of the segment's end point and the segment's origin point \textit{i.e.,} the elevation change of a trip is computed only counting uphills. We retrieve the elevation of 
a given GPS point using Microsoft Virtual Earth (currently Bing Maps Platform API). Formally, the rate of uphill elevation change for a trip is calculated as
$\frac{\text{(uphill elevation change (meter))}}{\text{trip duration (min)}}$ with uphill elevation change for a trip computed as the sum of the positive elevation changes across all segments in a trip: $\frac{\sum_{s=1}^{n}\text{uphill elevation change(s)}}{n}$ with $n$ being the total segments in the trip with a positive elevation change.}

To answer the research questions, we first we divided all the trips into pedelec and conventional bicycle trips and retrieved trip length, trip duration, speed, and rate of uphill elevation change values for all the trips in each group. 
\textcolor{black}{For each of these trip characteristics, we run a Welch's t-test between the value distributions 
for pedelecs and the values for conventional bicycles, to assess whether the usage differences between the two were statistically significantly different. Null hypotheses were rejected at a significance level of $1\%$ \textit{i.e.,}  $pvalue=0.01$ (strong evidence) or $5\%$ \textit{i.e.,} $pvalue=0.05$ (moderate evidence). Unlike the Student’s t-test, the Welch's t-test allows to compare populations that have unequal variances and unequal sample sizes. }

\textcolor{black}{Next, to analyze the difference in terms of types of origin and destination locations visited during the cycling trips,
we grouped trips by their start and end stations and compared the volume of trips across origin-destination pairs for pedelecs and bicycles. In addition, and to infer potential trip purpose, we characterize the types of stations visited using the authors' knowledge of the city as well as the city zoning codes. Specifically, we assign each bike station to the zoning code associated with the region where the station is located. For example, a station located in Richmond's Barton Heights is labeled as residential given that Barton Heights is a residential area according to Richmond's zoning codes. Once all stations have been labeled according to zoning codes, we calculate the volumes of trips between different pairs of origin-destination zoning codes. }

\textcolor{black}{We use the zoning codes from Richmond City, which are publicly available\footnote{The official interactive map of Richmond, which contains zoning information, can be found at \url{https://www.arcgis.com/apps/webappviewer/index.html?id=0077af87647f42cfbe5f4b2c6d05e6e2}. A detailed description of each zoning code can be found in Chapter 30 - zoning of \textit{Code of the city of Richmond} at \url{https://library.municode.com/va/richmond/codes/code_of_ordinances?nodeId=CH30ZO_ARTIVDIRE}}. Given the large number of zoning codes - for example, there are more than ten different codes to characterize residential areas - and to be able to carry out an interpretable analysis, we group the zoning codes based on general purpose. All codes starting with R and a number - that represent different types of residential areas - are classified as "residential". Codes RO (Residential-Office), RF (Riverfront), UB (Urban business) and DCC (Downtown) are categorized as "mixed use"; and codes B (Business), M (Industrial district) and RP (Research Park) are categorized as "business". All other codes are not very frequent, and we group them into the category "other".}

\textcolor{black}{Finally, to look at the differences between types of roads used for pedelecs and bicycles, we mapped the GPS trajectories of each trip using Mapbox’s Map Matching API as explained earlier; and analyzed the types of roads used by pedelec and bicycles, with a focus on major roads, minor roads and cycleways. To carry out this analysis, we also used Welch's t-tests to assess whether the use differences across road types where statistically significantly different at a significance level 
of $1\%$ \textit{i.e.,}  $pvalue=0.01$ (strong evidence) or
of $5\%$ \textit{i.e.,}  $pvalue=0.05$ (moderate evidence).}

\section{Results}
\subsection{Spatio-Temporal Analysis}

\iffalse
Using all trips, and their unique bike identifiers, we extracted the  
%we gathered the number of unique bikes used in a given month to estimate the 
number of bikes and pedelecs available. In April, about 25\% of the fleet was pedelec bikes and by December approximately 65\% (as we showed in Figure \ref{fig:13}). Figure \ref{fig:12} shows the rate of trips made per available bike. T-tests results showed that the mean number of trips made per bike available was significantly more (~3.2x) for pedelecs compared to bikes (p-value=0.004). We also notice that the system removes a large number of bikes over the time, and overall fleet size is reduced.

\begin{figure}[!ht]
  \centering
  \includegraphics[width=0.6\textwidth]{figures/figure12.png}
  \caption{ Number of Trips Made Per Available Bike or Pedelec extracted from the first week of each month.}\label{fig:12}
\end{figure}
\fi

\begin{table}[]
\centering
\caption{\textcolor{black}{Use Differences between Pedelecs and Bikes Considering All Trips. $pvalue=0.01$ refers to strong evidence (1\% significance level), $5\%$  or $pvalue=0.05$ reflects moderate evidence, and N.S. points to not statistically significant.} \textcolor{black}{Total speed calculations include stopping times.}}
\begin{tabular}{p{5cm}p{3cm}p{3cm}l}
Variable                         & PedelecMean (N=2259) & BicycleMean (N=1263) & p-value  \\\hline
Total Distance (km)            & 4.68                  & 3.85                                                                  & $< 0.01$ \\ %6.51E-11 \\
Rate of Elevation Change (m/m) & 0.00762               & 0.00876                                                               & $< 0.01$ \\ %6.81E-08 \\
Trip Time (min)                  & 30.7                  & 35.3                                                                  & $< 0.01$ \\ %2.12E-05 \\
Total Speed (kph)                & 10.5                   & 7.56                                                                   & $< 0.01$ %6.04E-93
\end{tabular}
\label{tab:4}
\end{table}

%We first analyzed the differences in the use of pedelecs and non-electric bicycles in Richmond, Virginia, for the period %under analysis (nine months). For that purpose, we divided all the trips into pedelec and bicycle trips and retrieved %trip length, trip duration, speed, and elevation values for all the trips in each group. Elevation was computed only %counting uphills, although the analysis with uphill and downhill values gave similar results. To evaluate if the %differences between these use variables across pedelec and bicycle trips were statistically significant, we ran t-tests %between each pair of distributions. 

We first analyzed the differences between pedelec and bicycle trips in terms of trip length, trip duration, speed and rate of uphill elevation change.
%$\frac{\text{(uphill elevation change (meter))}}{\text{trip duration (min)}}$
%We compute the rate of elevation change by dividing the absolute elevation change with the trip duration, defined as $\frac{\text{(abs(elevation change (meter))}}{\text{trip duration (min)}}$.
Table \ref{tab:4} shows the mean values for each of the use variables and \textcolor{black}{their corresponding statistical significance p-value.} 
%\textcolor{blue}{indicates whether the differences between the values associated with pedelec and bicycle are significant.} 
We can observe that pedelecs are associated with longer trip distances, shorter trips times, and higher speeds. These results highlight the fact that pedelecs are faster (10.5 kph vs. 7.56 kph) and as a result trip times tend to be shorter. Interestingly, we also observe that trip distances are longer, pointing to pedelecs being used for longer trips than bicycles. 
\textcolor{black}{These three findings around trip times, distances and speeds are similar to prior work discussed in the related work~\cite{langford2013north,fyhri2015effects,cherry2016dynamics}.
On the other hand, the rate of uphill elevation rate is lower for pedelecs than for bikes. 
\textcolor{black}{This finding is counterintuitive since we were expecting pedelecs to be used to overcome higher elevations. However, we posit that this finding could be related to the fact that Richmond, VA, is relatively flat and as prior work has shown, significant route shifts do not occur with small slope changes (of up to 2\% )  \cite{broach2012cyclists}}.}

\textcolor{black}{The speeds reported are lower compared to previous studies. We posit that this could be due to how RVA calculated the speeds: dividing the total distance by the total trip duration, which may have included times when the pedelecs or bicycles were stationary, thus lowering the overall trip speed. Secondly, considering these were the inaugural pedelecs introduced into Richmond's bike share system, it is plausible that cyclists exercised caution while using them, motivated by concerns over potential accidents involving pedelecs. This caution could have led to initially lower speeds, which might have increased as users became more accustomed to the pedelecs.}
%However, due to the unavailability of more recent speed data from RVA, we cannot substantiate the latter hypothesis.}

\begin{table}[]
\caption{\textcolor{black}{Use differences between Pedelecs and Bicycles Considering AM Commuting Trips. $pvalue=0.01$ refers to strong evidence (1\% significance level), $5\%$  or $pvalue=0.05$ reflects moderate evidence, and N.S. points to not statistically significant.}\textcolor{black}{Total speed calculations include stopping times.}}\label{tab:7}
\centering
\begin{tabular}{p{5cm}p{3cm}p{3cm}l}
Variable                         & PedelecMean (N=85) & BikeMean (N=24) & p-value  \\\hline
Total Distance (km)            & 4.84                                                         & 2.95                                                              & $ < 0.01$ \\ %1.16E-03 \\
Rate of Uphill Elevation Change (m/m) & 0.00592                                                             & 0.00469                                                           & N.S. \\ %6.37E-02 \\
Trip Time (min)                  & 31.0                                                                & 24.2                                                              & N.S.\\ %1.40E-01 \\
Total Speed (kph)                & 9.98                                                                 & 8.56                                                               & N.S.  %6.60E-02
\end{tabular}
\end{table}

\begin{table}[]
\centering
\caption{\textcolor{black}{Use Differences between Pedelecs and Bikes Considering Touring Trips. $pvalue=0.01$ refers to strong evidence (1\% significance level), $5\%$  or $pvalue=0.05$ reflects moderate evidence, and N.S. points to not statistically significant.}\textcolor{black}{Total speed calculations include stopping times.}}\label{tab:5}
\begin{tabular}{p{5cm}p{3cm}p{3cm}l}
Variable                         & PedelecMean (N=788) & BikeMean (N=607) & p-value  \\\hline
Total Distance (km)            & 6.20                                                                 & 4.72                                                               & $< 0.01$ \\ %6.34E-10 \\
Rate of Uphill Elevation Change (m/m) & 0.00811                                                              & 0.00906                                                            & $< 0.01$ \\ % 1.27E-03 \\
Trip Time (min)                  & 43.7                                                                 & 46.0                                                               & N.S \\% 1.87E-01 \\
Total Speed (kph)                & 8.53                                                                  & 6.44                                                                & $< 0.01$ %3.28E-26
\end{tabular}
\end{table}

\begin{table}[]
\centering
\caption{\textcolor{black}{Use differences between Pedelecs and Bicycles Considering O-D Trips. $pvalue=0.01$ refers to strong evidence (1\% significance level), $5\%$  or $pvalue=0.05$ reflects moderate evidence, and N.S. points to not statistically significant.}\textcolor{black}{Total speed calculations include stopping times.}}\label{tab:6}
\begin{tabular}{p{5cm}p{3cm}p{3cm}l}
Variable                         & PedelecMean (N=1471) & BikeMean (N=656) & p-value  \\\hline
Total Distance (km)            & 3.91                                                                  & 3.09                                                               & $< 0.01$ \\ %1.50E-09 \\
Rate of Uphill Elevation Change (m/m) & 0.00736                                                               & 0.00849                                                            & $< 0.01$ \\ %1.33E-04 \\
Trip Time (min)                  & 24.0                                                                  & 26.0                                                               & N.S. \\ % 9.71E-02 \\
Total Speed (kph)                & 11.4                                                                  & 8.53                                                                & $< 0.01$ \\ %4.22E-57
\end{tabular}
\end{table}

 \textcolor{black}{In an attempt to understand in more depth the differences between pedelec and bicycle, we defined three types of trips: touring (start and end points are the same), O-D trips (different start and end points) and AM commuting trips (trips between 6-10am during weekdays) and analyzed statistically significant differences between pedelec and bicycle trips using Welch's t-tests as described before. Table \ref{tab:7} through Table \ref{tab:6} show the results for this analysis. }

For AM commuting trips (see Table \ref{tab:7}), the only significant difference between pedelec and bikes was the total distance, with pedelecs making longer trips (4.84 km vs. 2.95 km). For touring trips (see Table \ref{tab:5}), total distance was longer (6.20 km vs. 4.72 km) and trip speeds higher (8.53 kph vs 6.44 kph) for pedelecs compared to bicycles; however, there was no significant difference in travel time. 
On the other hand, while O-D trips had a much lower average distance (see Table \ref{tab:6}), there was still a significant difference between trip distance for pedelecs (3.91 km) versus bikes (3.09 km). As with touring trips, the difference in travel time was not significant.
\textcolor{black}{Finally, pedelec trips were associated with significantly lower rates of uphill elevation change, except for AM commuting trips where elevation differences between pedelecs and conventional bicycles were not significant, possibly pointing to convenience and speed when going to work.}
\textcolor{black}{The significantly higher rates of uphill elevation change for bikes in touring trips might point to an exercise purpose.}

\subsection{\textcolor{black}{Trip Analysis by Types of Origin and Destination Locations Visited}}

\begin{figure}
\centering
\begin{subfigure}{.49\textwidth}
  \centering
  \includegraphics[width=\textwidth]{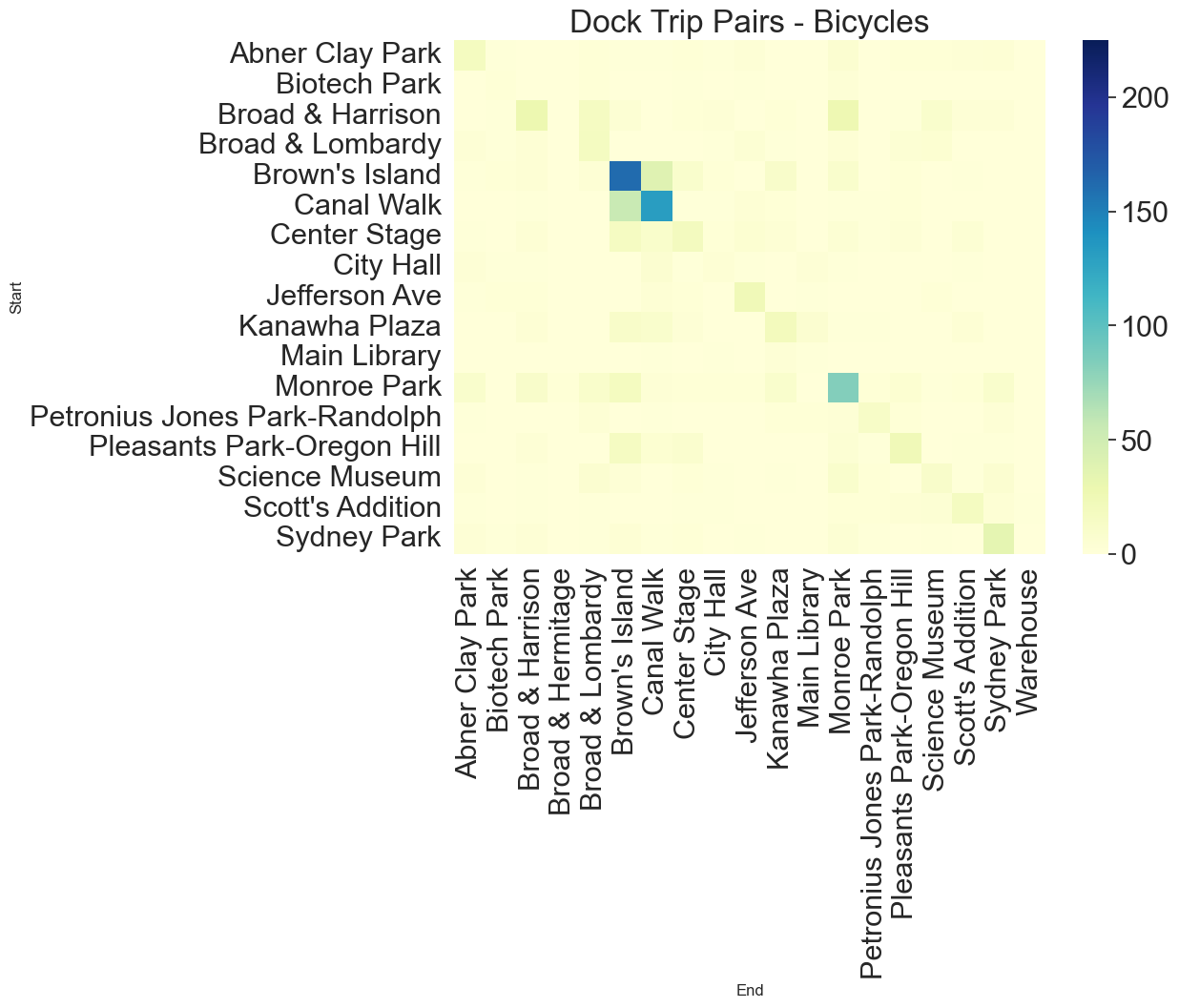}
  \caption{Start and End Trip Pairs for Bicycles.}
  \label{fig:14}
\end{subfigure}
\begin{subfigure}{.49\textwidth}
  \centering
  \includegraphics[width=\textwidth]{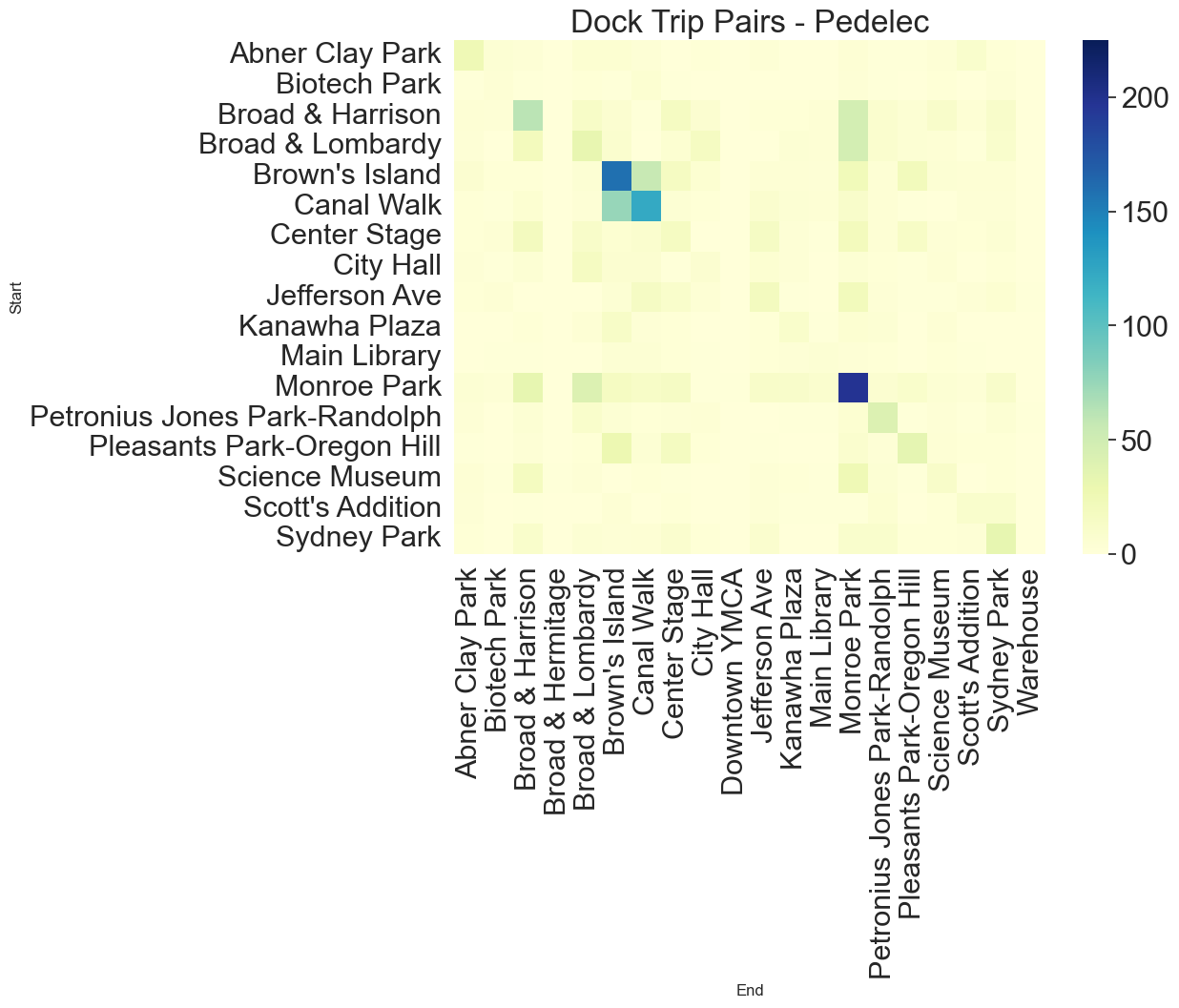}
  \caption{Start and End Trip Pairs for Pedelecs.}
  \label{fig:15}
\end{subfigure}
\begin{subfigure}{.49\textwidth}
  \centering
  \includegraphics[width=\textwidth]{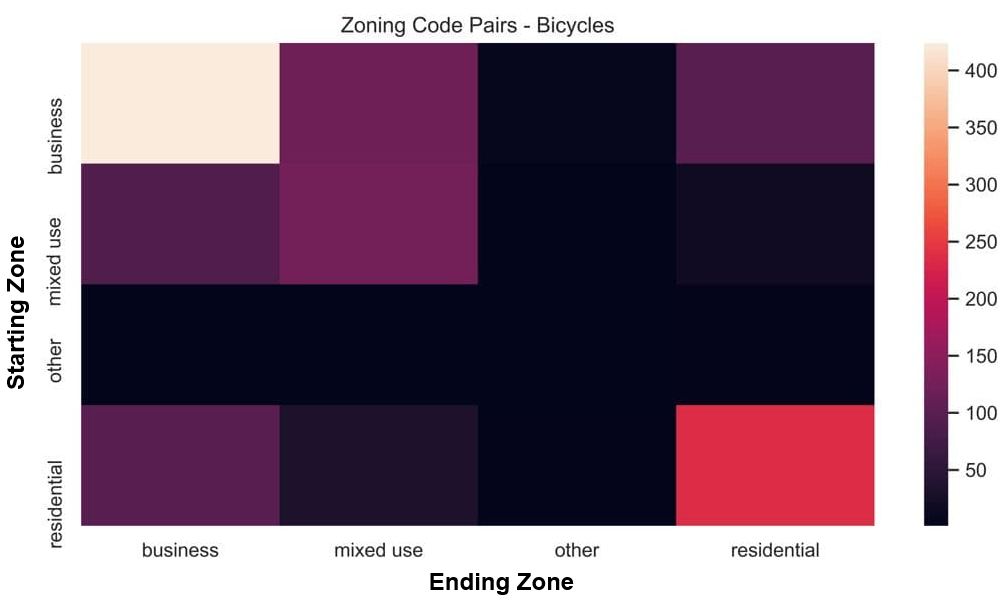}
  \caption{Start-End Zoning Code Pairs for Bikes.}
  \label{fig:16}
\end{subfigure}
\begin{subfigure}{.49\textwidth}
  \centering
  \includegraphics[width=\textwidth]{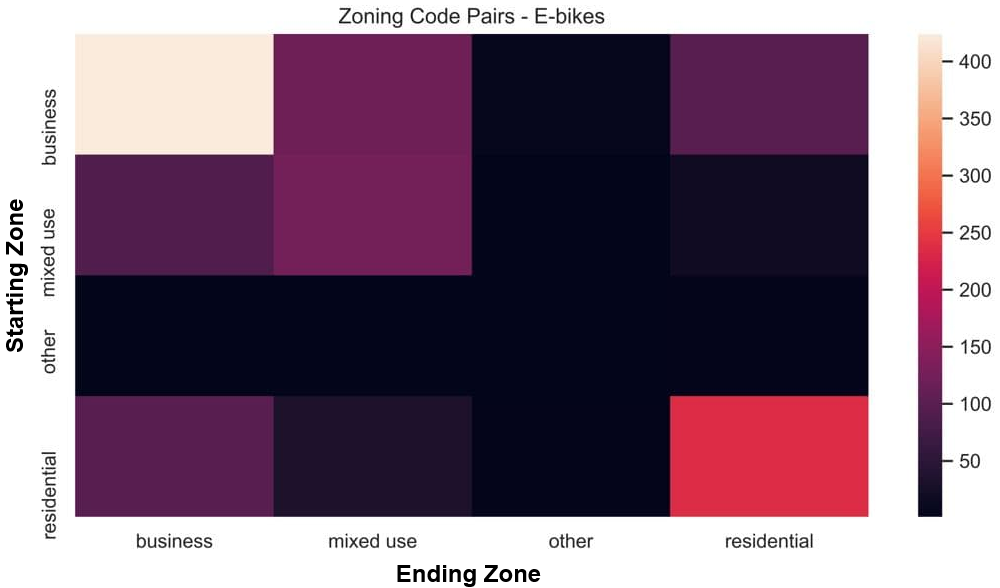}
  \caption{Start-End Zoning Code Pairs for Pedelecs.}
  \label{fig:17}
\end{subfigure}
\caption{Trip analysis for Bikes and Pedelecs.}
\label{fig:test}
\end{figure}

\textcolor{black}{To analyze the difference in terms of types of origin and destination locations visited during the cycling trips, we grouped trips by their start and end stations and compared the number of trips across destination pairs for pedelecs and bicycles. To infer potential trip purpose, we characterize the types of stations visited using the authors' knowledge of the city and the city zoning codes. Figure \ref{fig:14} and Figure \ref{fig:15} show the total number of trips for each pair of start and end stations, respectively. The diagonal displays the number of touring trips (same origin and destination stations), while the others represent O-D trips (with different origin and destination stations.} 

\textcolor{black}{Consistent with the previous analysis, these plots show that the most frequent trips for both pedelecs and bicycles are trips that start and end at Brown’s Island/Canal Walk and at Monroe Park, well known green areas in Richmond, pointing to exercise use since these are touring trips (same origin and destination station) that go through green areas. The second most popular trips are those that start and end at Broad and Harrison streets, a downtown Richmond location area full of retail businesses, pointing to secondary purposes potentially related to shopping or recreational activities.} It is also important to highlight that a chi-square test between the pedelec and the bicycle distributions revealed that both distributions were different \textcolor{black}{(at 1\% significance level).}

In an attempt to better understand the types of locations visited, we also grouped trips by the type of origin and destination zoning code. We considered business, mixed use, residential and other uses (see Section 3.5 for details). 
 Figure \ref{fig:16} and Figure \ref{fig:17} show the volumes of trips for each zoning code pair for pedelecs and bicycles, respectively. We can observe that the plots show extremely similar trends with a large number of trips staying within either business or residential locations. We also observe considerable volumes of trips between mixed use and residential zoning codes, followed by business and residential. We posit that while the “other” trips might be representing exercise-type trips, the ones related to business and mixed use might reflect either shopping or other recreational activities, as discussed earlier.

\subsection{Roadway Use}
\begin{figure}
\centering
\begin{subfigure}{.49\textwidth}
  \centering
  \includegraphics[width=\textwidth]{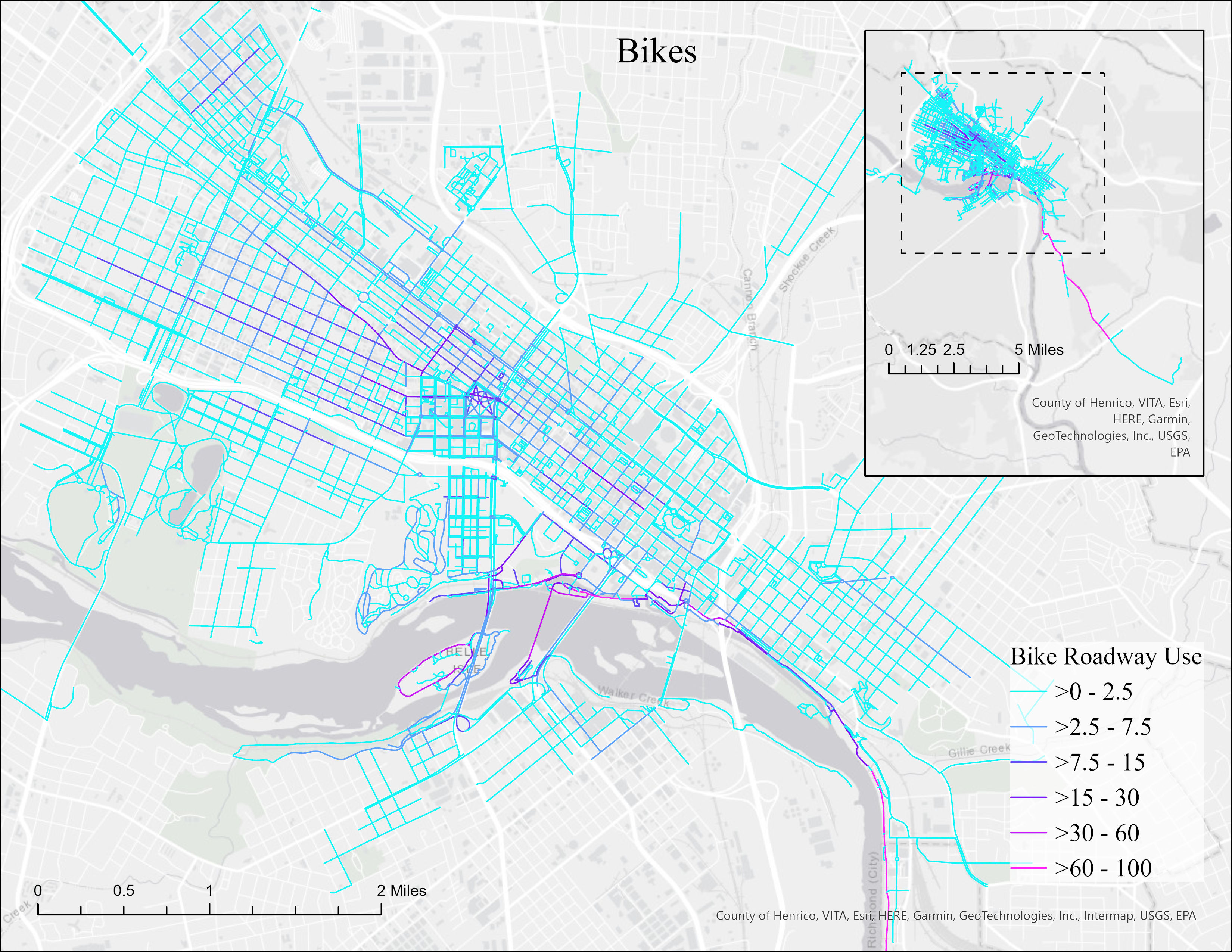}
  \caption{Bikes}
  \label{fig:18}
\end{subfigure}
\begin{subfigure}{.49\textwidth}
  \centering
  \includegraphics[width=\textwidth]{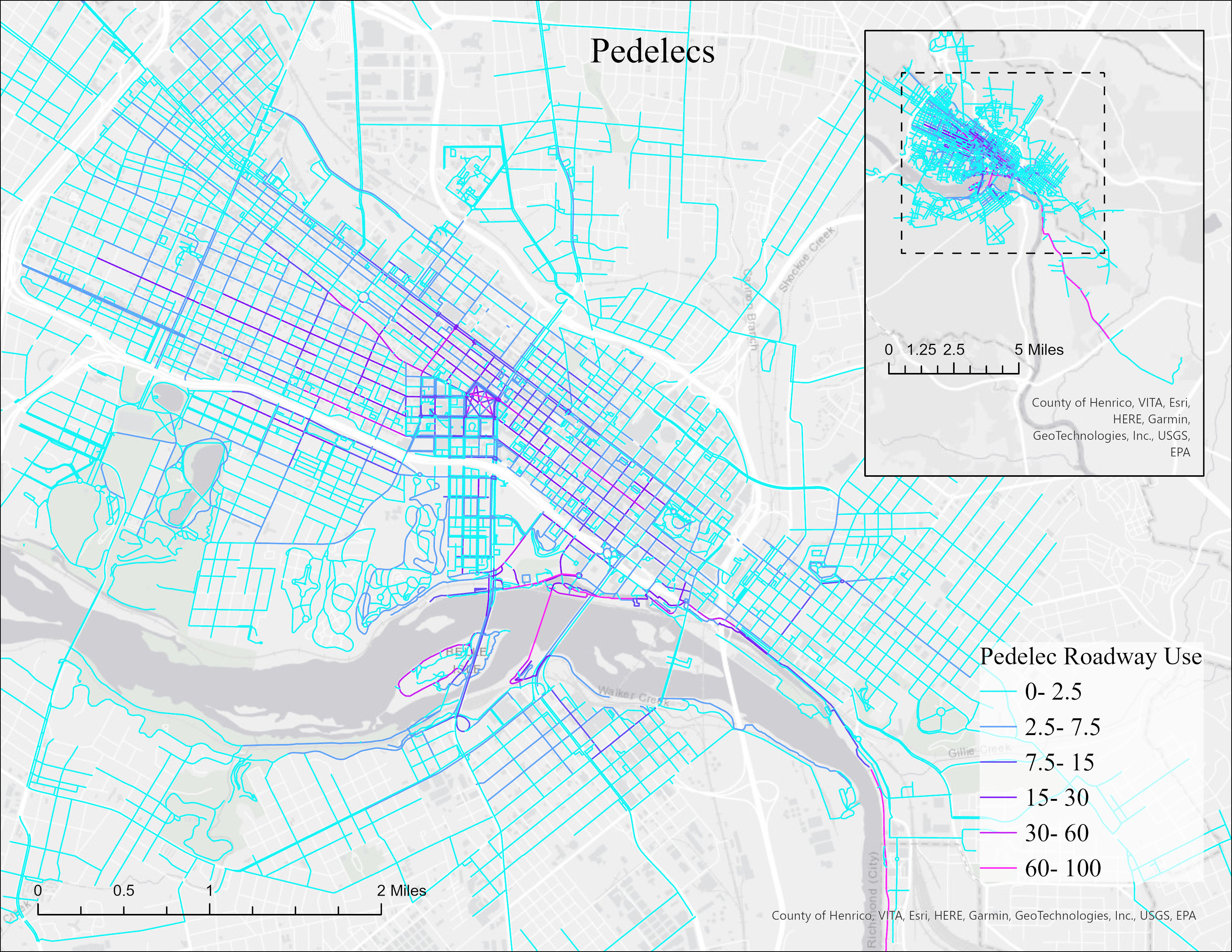}
  \caption{Pedelecs}
  \label{fig:19}
\end{subfigure}
\caption{Normalized roadway usage by bike type. \textcolor{black}{The basemaps were generated with Esri ArcGIS and the following sources: Esri, HERE, Garmin, GeoTechnologies, Inc., USGS and EPA.}}
\label{fig:test}
\end{figure}

%To look at the routing differences between pedelecs and bicycless, we mapped the GPS trajectories of each trip %using Mapbox’s Map Matching API as explained earlier; we break the snapped roads into segments and query each segment in %Open Street Maps (OSM) to identify the type of road. 
Figure \ref{fig:18} and Figure \ref{fig:19} show the roadway segments used by pedelecs and bikes, respectively. The total number of trips that occurred on each segment was normalized by the number of trips on the segment with the maximum number of trips multiplied by 100. Thus, the road in pink has a frequency of 60\%-100\% of the road with the maximum number of trips. Pedelecs were used farther outside of the city than bikes. Additionally, pedelecs were frequently used in the downtown core where most RVA bike share stations are located. Both bikes and pedelecs were frequently used along the riverfront, Belle Isle, and the bike trail that runs south of downtown starting along the riverfront.

\textcolor{black}{To analyze the types of roads traveled, we assigned each street segment in a trip to a label characterizing the type of road. For analytical purposes, we considered three types of roads: major roads, minor roads and roads with cycleways (see Section 3.5 for details).}
%\textcolor{blue}{
%We used the road typology definitions provided by Open Street Maps (OSM) to define these three types and defined as major roads roads with the OSM labels "motorway", "trunk", "primary", "secondary” and "tertiary"; as minor roads, road segments with the OSM labels “road”, “residential”, “living_street”, “service”, and “pedestrian”; and, road segments with the OSM label “cycleway” were considered cycleways. }
\textcolor{black}{For each type of segment, we report two metrics (i) the average number of kilometers traveled, computed across all trips for pedelecs and bicycles, (ii) the average percentage of kilometers traveled during each trip by pedelec or bicycle, and (iii) the significance of the Welch's t-test that compares whether there exist statistically significant differences between the two metrics across pedelecs and bikes.  
We then ran Welch's t-tests to understand whether the differences between pedelec and bicycle use with respect to road usage was statistically significantly different at a significance level of 1\% or 5\%. }

Tables \ref{tab:9} and Table \ref{tab:8} summarize our results. Next, we discuss the main significant results, all at a $1\%$ significance level. 
%for three different types of segments: major roads, minor roads, and roads with cycleways. 
%the p-value that evaluates whether the reported differences are statistically significant (p<0.05 shows that the difference in mean is significant).
\textcolor{black}{The tables show that pedelecs are used for longer distances on major roads compared to bicycles, averaging 2.06 km versus 1.35 km for bicycles. Additionally, a larger portion of the trip is on major roads for pedelecs, at 45\%, compared to 36\% for bicycles. On the other hand, pedelecs are associated with smaller trip portions on minor roads (55\% vs. 64\%). Looking into roads with cycleways, Table \ref{tab:8} shows that pedelec users used roads with cycleways for a greater proportion of their trip than conventional bicycle users, and Table \ref{tab:9} highlights pedelecs are used for longer distances on cycleways than bicycles, averaging 1.61 km versus 1.13 km. }

\begin{table}[]
  \centering
  \caption{\textcolor{black}{Welch's t-test for average kilometers traveled across trips by pedelec or bike on select roadway types. $pvalue=0.01$ refers to strong evidence (1\% significance level), $5\%$  or $pvalue=0.05$ reflects moderate evidence, and N.S. points to not statistically significant.}}\label{tab:9}
\begin{tabular}{llll}
Mean km & Major Roads & Minor Roads & Cycleways \\\hline
Bike       & 1.35        & 2.64        & 1.13      \\
Pedelec    & 2.06        & 2.78        & 1.61      \\
p-value    & $< 0.01$    & N.S.    &$< 0.01$  
%p-value    & 5.17E-19     & 1.73E-01    & 3.19E-09 
\end{tabular}
\end{table}

\begin{table}[]
  \centering
  \caption{\textcolor{black}{Welch's t-test for average percentage of road use per trip by Pedelec or Bike on select roadway types. $pvalue=0.01$ refers to strong evidence (1\% significance level), $5\%$  or $pvalue=0.05$ reflects moderate evidence, and N.S. points to not statistically significant.}}\label{tab:8}
\begin{tabular}{llll}
Mean percentage & Major Roads & Minor Roads & Cycleways  \\\hline
Bike                                    & 36.00\%                             & 64.00\%                             & 28.21\%                           \\
Pedelec                                 & 45.16\%                             & 54.84\%                             & 35.22\%                             \\
p-value                                 & $< 0.01$                             & $< 0.01$                             & $< 0.01$   
%p-value                                 & 7.52E-17                            & 7.52E-17                            & 4.26E-12
\end{tabular}
\end{table}

\section{Conclusions}
\textcolor{black}{This work has presented a comprehensive analysis of the similarities and differences between pedelec and conventional bike trips in a bike share system in Richmond, Virginia.
Our results have shown that pedelecs are generally associated with longer trip distances, shorter trips times, higher speeds and lower rates of uphill elevation change. These results were similar across types of trips: touring, O-D and AM commuting, with the exception of AM commuting trips where elevations were higher, possibly pointing to convenience and speed when going to work. The study area is relatively flat; thus, future work should consider the impact of elevation in an area with hillier terrain. }

Pedelecs were also found to be associated with higher average \textcolor{blue}{trip distances on major roads and cycleways than bikes, and with lower percentages of minor road use than bikes.} \textcolor{black}{Origin-destination analysis on pairs of stations has shown popular pairs mostly associated with exercise (green areas) and recreational activities both for pedelec and bicycles. }
Future work will explore the similarities and differences between existing and new users, potentially attracted by the introduction of e-bikes in the shared system.

\section{Funding}
This work was supported by the by the National Science Foundation [NSF-1951924, NSF-1750102]; the Federal Highway Administration and the Urban Mobility \& Equity Center.

\bibliographystyle{elsarticle-num-names} 
\bibliography{trb_template}
%\bibliography{main}

\end{document}